\documentclass[amsmath,amssymb,aps,pra,reprint]{revtex4-2}
\usepackage{graphicx}
\usepackage[]{hyperref}

\begin{document}

\title{Griffiths physics in an ultracold Bose gas}

\author{M. E. W. Reed}
\author{Z. S. Smith}
\author{Aftaab Dewan}
\author{S. L. Rolston}
\email{rolston@umd.edu}
\affiliation{Joint Quantum Institute, University of Maryland and National Institute of Standards and Technology, College Park, Maryland 20742, USA}

\date{\today}

\begin{abstract}
Coupled \emph{XY} model systems consisting of three-dimensional (3D) systems with disordered interlayer physics are of significant  theoretical interest. We realize a set of coupled quasi-2D layers of $^{87}$Rb in the presence of disordered interlayer coupling. This is achieved with our high bandwidth arbitrary optical lattice to obviate restrictions on the dimensionality of disorder with speckle-generated optical fields. We identify phase crossover regions compatible with the existence of a pair of intermediate Griffiths phases between a thermal state and the emergence of bulk 3D superfluidity.
\end{abstract}

\maketitle

\section{Introduction}
The intersection of dimensionality and disorder is a rich area of study in condensed-matter physics.  The precise control of disorder available with optical potentials enables the realization of well characterized disordered systems with quantum degenerate atomic gases. Optical speckle has been used to generate disorder for one-, two-, and three-dimensional systems \cite{1DLoc,BourdelDis,DeMarcoLoc}, exhibiting Anderson localization, a disordered Berzinskii-Kosterlitz-Thouless (BKT) transition \cite{BeelerDis}, mobility edges in three dimensions \cite{Inguscio3D}, and emergence of a Bose glass \cite{DeMarcoGlass}. Similarly, quasidisorder provided by incommensurate lattices has been used to realize the Aubry-Andr\'e model in the presence of interactions \cite{Ing1D}, the role of quasi-disorder in adiabaticity \cite{EmilyDis}, to study many-body localization \cite{MBL}, and has been contrasted  against uncorrelated disorder in transport \cite{GadwayDisorder}.

While two-dimensional (2D) Bose gases in isolation are well described by BKT physics \cite{2DBoseReview} (approximated by the \emph{XY} model), recent theoretical work \cite{DemlerSliding,VojtaSliding,LFSliding} suggests rich behavior when layers of such systems (including stacks of 2D superfluids, cuprate superconductors, and planar magnets) are randomly coupled to one another.
They are predicted  to exhibit Griffiths physics \cite{MagGrifSum,Griffiths2,OrigGriffiths} when the interlayer couplings or layer thicknesses are subject to uncorrelated disorder. A pair of phases of matter emerge as the temperature is lowered from a nondegenerate state to bulk 3D order (Bose-Einstein condensate, magnetization, or superconduction).  Each intermediate phase is a Griffiths phase,  with properties dominated by the most extreme local deviation in the disordered system. The first is an anomalous Griffiths phase, a class of sliding phase \cite{SlidingDNA} that exhibits 2D order (superfluidity, magnetic susceptibility, or superconduction). The second is a semiordered Griffiths phase where order appears in the third dimension. (Magnetic Griffiths phases have recently been observed in bulk metal alloy systems \cite{MagGrifOne,MagGrifTwo}, but not in anisotropic systems.)

References \cite{VojtaSliding,DemlerSliding,LFSliding} employ the phase stiffness $\rho_s^{x}$ along a direction $e_x$ to characterize the Griffiths phases. Phase stiffness is a measure of the energy required to impose a phase difference $\phi_{x}$ between two ends of a finite system. As a function of  system size $L$ and system energy $E$, the phase stiffness is defined as
\begin{equation}
\rho_s^{x}=\frac{1}{L}\frac{d^2E(\phi_{x})}{d\phi^2_{x}}.
\end{equation}
While a computationally convenient parameter, it is not easily measurable in experimental atomic systems. Although it is possible to measure the critical velocities of superfluids \cite{DalibardSFV}, these disordered systems are predicted to have very small superfluid fractions and critical velocities over much of the phase diagram.  Instead we use an analysis of the fluctuations in large data sets of time-of-flight (TOF) momentum distributions to gain information about the phases as a function of temperature and lattice depth.

We associate the appearance of a Thomas-Fermi-like distribution in the two in-plane dimensions ($p_{\parallel}$) paired with a thermal distribution in the third ($p_{\bot}$) with a 2D superfluid transition at $\sim$200 nK.  As we decrease the temperature we observe the emergence of discrete modes in fluctuation correlations along $p_{\bot}$ \cite{MatheyCorr}, which may characterize the expanding length-scales of superfluid puddles in the anomalous Griffiths phase. Finally we interpret the suppression of zero momentum atomic density fluctuations at our lowest temperatures with the Bose statistics of macroscopically occupied non-local state, an observation consistent with 3D superfluidity.

\section{Experimental Approach}
We study this system using optical potentials generated for a cloud of ultracold $^{87}$Rb.
After initial loading and cooling, the atoms are held in vacuum by a pair of optical tweezers.
There they are subjected to a disordered 1D optical potential from an optical lattice generation system.
They are further cooled via optical evaporation to temperatures from  10 to 250  nK.
This temperature range spans both thermal and degenerate states of the atomic cloud.
After some evolution time in the lattice, we release the atoms from all optical fields to expand ballistically.
They form a position distribution determined by their momentum distribution at the time of release.
This is repeated many times, allowing us to study statistics of the measured momentum distributions.
Each part of this process is discussed in more detail below.

We create degenerate gases of $^{87}$Rb using a hybrid magnetic trap-optical dipole trap \cite{Chamber}, which then loads into a dipole trap whose waist is translated 30 cm into a science chamber.  The atoms are finally loaded into a crossed 1550-nm dipole trap and disordered optical lattice, and evaporatively cooled in the $F=1$, $m_F=-1$ state.   Final trapping frequencies ($\nu_{\parallel}$, $\nu_{\bot}$) are  $E/2\pi\hbar$ = (100.0, 20.0) Hz at our lowest temperatures and (126.5, 25.3) Hz at our highest.

\begin{figure}
 \includegraphics[scale=.28]{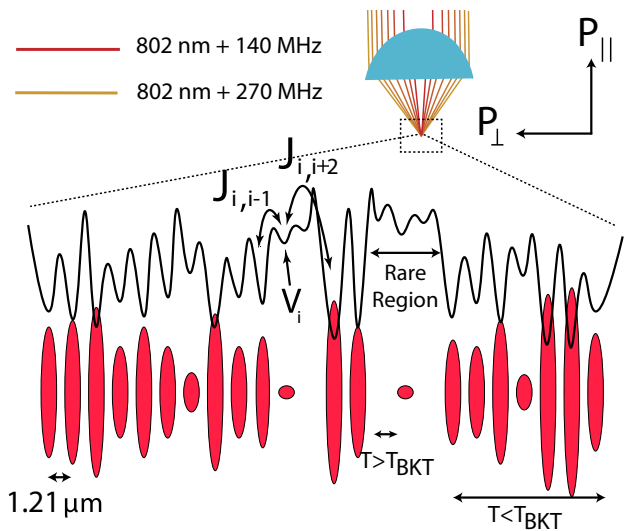}%
 \caption{\label{App}We engineer a stack of superfluid pancakes with a set of phase-stabilized shallow-angle lattices formed by our HiBAL. Its set of phase-stabilized parallel beams is focused through an aspheric lens to create the 1D disordered optical potential. The local minima have different depths, which generates a local phase space density for each pancake. Interplane hopping $J_{i,j}$ is subject to disorder as well, matching our system most closely with Ref. \cite{DemlerSliding}.}
 \end{figure}

Evaporation begins with a nondegenerate cloud in the presence of our high bandwidth arbitrary lattice (HiBAL) (see Fig. \ref{App}).
(Disorder generated using optical speckle necessarily includes variations along the propagation direction
\footnote{For 1D speckle disorder to be feasible, the Rayleigh range must be much larger than the cloud size. Our clouds are on the order of $\approx$(10--100) $\mu m$, whereas for a speckle with features and wavelength $\approx1$ $\mu m$, the Rayleigh range ($z_R = \pi w^2/\lambda$) is 4 $\mu m$.}, so the HiBAL was developed to provide the desired potentials.)
We Fourier synthesize a potential $U=\sum_j A_j \text{sin}(k_j x+\phi_j)$ from multiple optical lattices spanning two octaves of spatial frequencies.  We produce a time-averaged disordered 1D optical potential using  802-nm light  with a set of seven shallow angle lattices of incommensurate period ranging from 1.21 to 4.70 $\mu$m. The 1.21-$\mu$m (base) lattice is phase stabilized to an rms of 6(1) mrad with a Mach-Zehnder interferometer and a piezoelectric transducer-mounted mirror. We confirm the costability of the other optical lattices to our base lattice and our atoms through our diffraction limited 0.5-NA, $22.8\times$ microscope to 19(3) mrad rms. Each lattice pulse is 300 ns in duration. The longest delay between base lattice pulses is 600 ns, and the pulse sequence repetition rate is 3 $\mu$s. We use two disordered lattice depths, labeled by the base lattice depth (2.8 and 5.6 kHz).  The summed average depth of all seven lattices is 3 times deeper, with an rms deviation equal to the base lattice depth.  Our deeper base lattice depth  corresponds to 14 effective recoil energies ($E_R/2\pi\hbar=0.4$ kHz). The rapid pulse rates of our lattices compared to their energy gaps puts us well within the time-averaged potential limit. We observe no differences in heating or TOF images between constant and stroboscopic lattices, thus do not need to consider Floquet physics effects.  Shot-to-shot variation in the position of our harmonic 1550-nm trap is sub-$\mu$m, and the fastest systematic drift during evaporation is 6 $\mu$m/s, which corresponds to energies too small to drive excitations.

The quasi-2D regime in cold gases is  reached when the fractional occupation of out-of-plane excited states in an effective, local harmonic oscillator is small. This occurs when the thermal energy scale $k_B T$ is smaller than the harmonic oscillator level spacing $\hbar \omega$,
or alternatively when the thermal deBroglie wavelength $\lambda_{dB}$ is larger than the spatial extent $\sqrt{2 \pi}a_z$ where $a_z$ is the harmonic oscillator length \cite{2DBoseReview}.
As described above, our disordered lattice has sites with varying depths. For our mean site depth of the shallow  lattice ($3\times 2.8$ kHz$ = 8.4$ kHz) together with the base lattice spacing,
we have $k_B T /(\hbar\omega_\bot) \leqslant 1 $ for $T \leqslant 250$ nK,  putting us in the quasi-2D regime throughout.

In this regime, our time-averaged Hamiltonian under the local density approximation is
\begin{gather}
\label{bsH}
\notag\hat{H} = \sum_{i} \left[ \frac{\hat{p}_{\parallel}^2}{2m} + V_i - \left( \sum_{j} \frac{1}{2} J_{i,j} \psi_i^\dag \psi_j \right)\right. \\
       \left. + \tilde{g}_i \psi_i \psi_i^\dag + \frac{1}{2} m \omega_i^2 \left( x^2 + y^2 \right) \vphantom{\left(\sum_j\frac{1}{2}\right)} \right],
\end{gather}
where $\hat{p}_{\parallel}^2/2m$ is the in-plane kinetic energy, $V_i$ is the lattice well depth, $J_{i,j}$ is the hopping strength between two lattice sites $i$ and $j$, $\tilde{g}_i$ is the scaled quasi-2D self-energy term for each pancake, $\omega_i$ is the  in-plane trapping frequency, which includes small variations due to the disorder \footnote{The out-of-plane momentum $p_\bot$ is contained in the $J_{i,j}$ and $\tilde{g}_i$ terms of the Hamiltonian.}.

We prepare gases ranging in temperature from 250 nK down to 25 nK in our lattice potentials.  The relative phases of our component lattices are arbitrarily tunable.  We chose two sets of relative phases to create two different optical potentials with similar statistics.  We observe no difference in the results between the two potentials, confirming that our atom cloud is large enough (greater than 70 pancakes) to average over the disorder. Each data set contains two different hold times, 200 ms and 4 s, which show no statistical difference, indicating equilibrium.

We levitate our atoms during 47 ms of TOF with the use of a gravity-canceling magnetic coil with a microsecond switching time to resolve the momentum distributions $n(p_{\parallel},p_{\bot})$. Gross-Pitaevskii simulations show that our distributions along the disordered direction, $p_{\bot}$, are minimally distorted by interactions during the expansion due to the rapid expansion of each atomic plane. We conclude that our TOF distributions faithfully represent the momentum distribution along the disordered direction.

\begin{figure}
\includegraphics{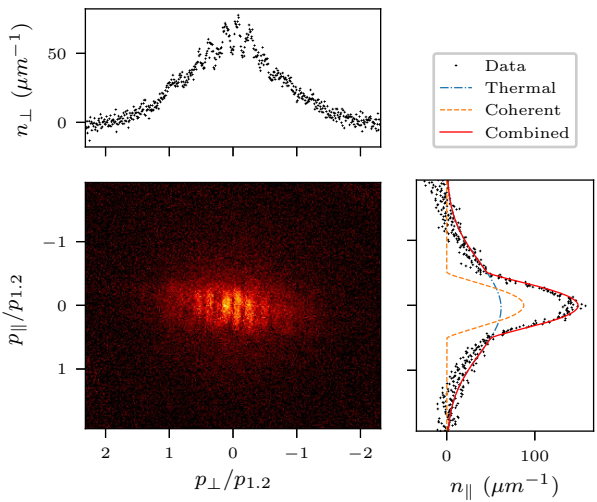}
\caption{\label{fig:eximg}
An example image from a 112-nK cloud in our 2.8 kHz lattice, shown together with the distributions $n_\perp$ and $n_\parallel$ for that image.
Momenta are given in units of the recoil momentum of our shortest period lattice, $p_{1.2}$.
The bimodal fit to $n_\parallel$ is displayed as a solid red line, with the thermal and coherent parts of the fit displayed separately as a blue dot dashed and orange dotted lines, respectively.}
\end{figure}

\begin{figure}
\includegraphics{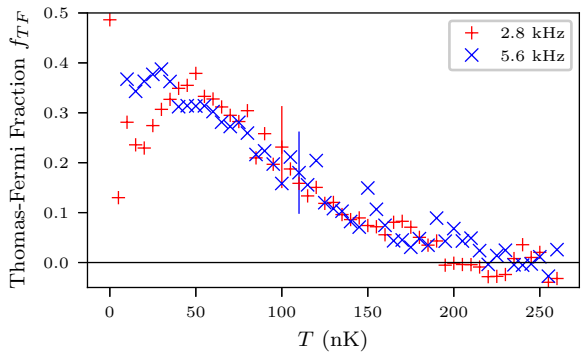}
\caption{\label{fig:cfrac}
The fraction of atoms in the Thomas-Fermi distribution, $f_\mathrm{TF}$, is plotted as a function of temperature.
The shallow lattice is plotted with red +'s, and the deep lattice with blue $\times$'s.
Error bars are the standard deviations of the fractions in each bin, and a typical bar is shown near the center for each lattice depth.
We remove a baseline value of 0.054, a fit artifact, from the Thomas-Fermi fraction.
$f_\mathrm{TF}$ is consistent with zero at high temperatures, but increases to finite values as early as 200 nK, establishing $\rho_s^{\parallel}>0$ below 200 nK.
}
\end{figure}

\section{Analysis}
The trapping energies along the disorder and in plane differ by over a factor of 85,
causing each direction to expand at very different rates.
The expansion from the tightest direction of confinement will quickly reduce the local density,
preventing any interactions that might have otherwise coupled $p_\parallel$ and $p_\bot$ during the ballistic expansion phase.
Therefore we treat their momentum distributions as separable. We focus on the in-plane and longitudinal momentum distributions, $n_{\parallel}(p)=\sum_{p_{\bot}}n(p_{\parallel},p_{\bot})$ and $n_{\bot}(p)=\sum_{p_{\parallel}}n(p_{\parallel},p_{\bot})$. At high temperatures $n(p_{\parallel})$  is a Gaussian. As the temperature drops a bimodal distribution emerges, a thermal distribution plus a Thomas-Fermi-like distribution as expected after expansion \cite{CastinDum}.
An example image is shown in Fig. \ref{fig:eximg}.
We apply a bimodal fit to $n(p_{\parallel})$ for every cloud, from which we can extract the total number in the distribution $N$ and the number in the Thomas-Fermi part $N_\mathrm{TF}$.
The Thomas-Fermi fraction (Fig. \ref{fig:cfrac}) is then computed as $f_\mathrm{TF}=N_\mathrm{TF}/N - 0.054$, corrected for an experimentally determined fit offset.
At low temperatures, the thermal distribution becomes comparable in size to the Thomas-Fermi part,
causing the fit to underestimate $f_\mathrm{TF}$.
Both lattice depths show the emergence of a Thomas-Fermi fraction in $n(p_{\parallel})$ near 200 nK.
This corresponds to a coherent part and finite in-plane phase stiffness.
We identify this as the onset of the BKT transition in isolated pancakes.
We expect little difference in the BKT transition temperature between the two lattice depths;
the band gap and the compression of the planes scale weakly with lattice depth in our relatively shallow lattices.

\begin{figure}
\includegraphics[]{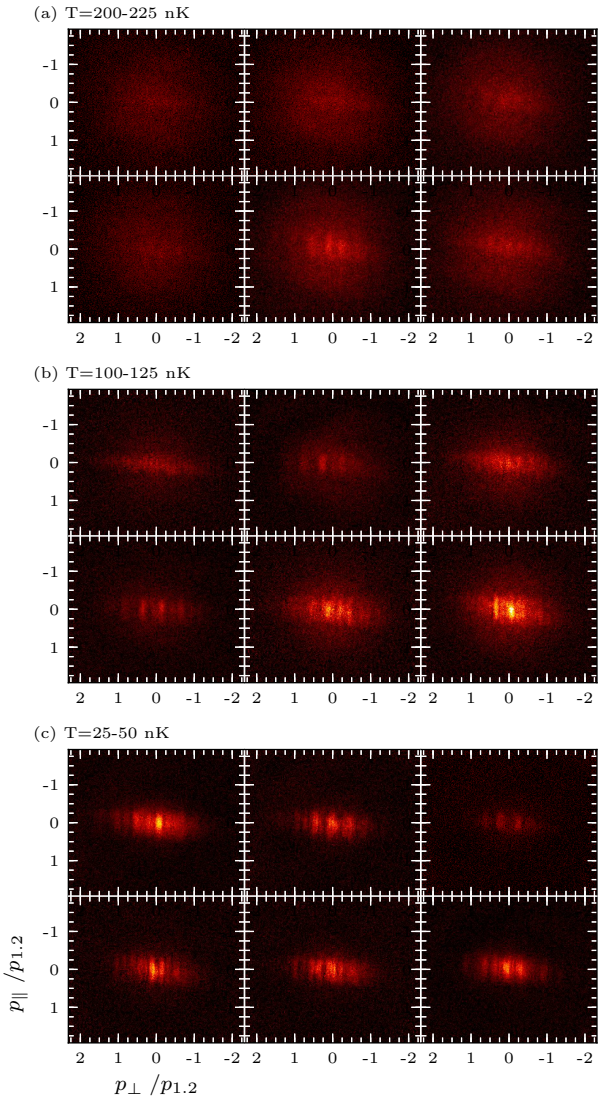}
\caption{\label{fig:zoo}
A random sampling of images at a few temperature ranges in our 2.8 kHz lattice. Top: 200--225 nK; middle: 100--125 nK; bottom: 25--50 nK.
}
\end{figure}
\begin{figure}
\includegraphics{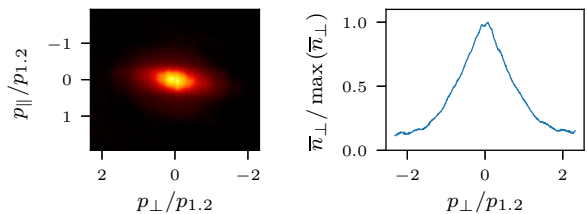}
\caption{\label{fig:avgimg}
The (left) average density profile for shots in the 2.8-kHz potential with temperatures between 25 and 50 nK,
and (right) $\overline{n}_\bot$ for the same conditions.
Neither shows any structure, despite the wide variation seen in Fig. \ref{fig:zoo}.
}
\end{figure}

\begin{figure}
\includegraphics{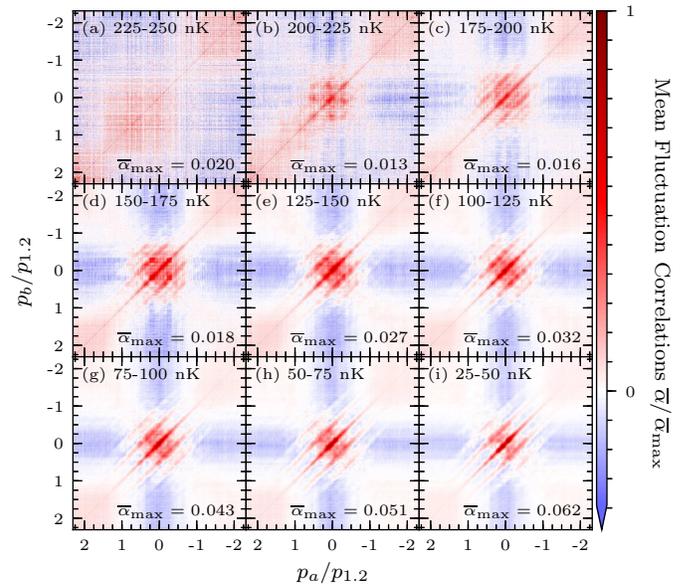}
\caption{\label{fig:alpha} We display the mean fluctuation corrections of the momentum, $\overline{\alpha}(p_a,p_b,T)$, at decreasing temperatures in panels (a)--(i), with the color scale in each panel normalized to the maximum $\overline{\alpha}_\mathrm{max}$.
The features at the edge of the panels are artifacts of our normalization scheme.
The temperature bins are 25 nK wide, and range monotonically from 225--250 nK in panel (a) to 25--50 nK in panel (i). Note the emergence of distinct momentum peaks in opposite momenta, which do not correspond to the sublattice recoil momenta.  The longest length scale resolved is 10.9 $\mu$m.}
\end{figure}

\begin{figure}
\includegraphics[]{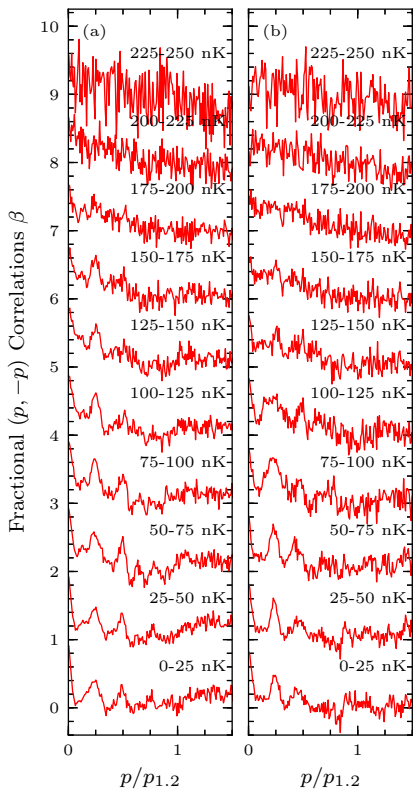}
\caption{\label{fig:beta}%
Normalized $p, -p$ correlations $\beta =2\overline{\alpha}(p,-p,T)/[\overline{\alpha}(p,p,T)+\overline{\alpha}(-p,-p,T)]$ for (a) the 2.8-kHz and (b) the deeper 5.6-kHz lattice.
Thermal correlations persist to lower temperatures in (b) than (a), despite similar coherent fraction as seen in Fig. \ref{fig:cfrac}.}
\end{figure}

At lower temperatures coherence between  planes emerges in the form of interference effects in $n_{\bot}(p)$.
Identically prepared clouds look radically different from one another in TOF (see Fig. \ref{fig:zoo} for example images).
The average of images at the same conditions is featureless, as shown in Fig. \ref{fig:avgimg}, so useful information is contained in the fluctuations.
Inspired by Feng \emph{et al.}'s success in identifying phases of matter in 1D Bose gases \cite{dpidpj}, which exhibit strong phase fluctuations, we apply a similar analysis to our data.
We normalize our distributions to $\sum_{p}n_{\bot}(p)=1$, sort them into 25-nK bins and calculate the average distributions $\overline{n}_{\bot}(p,T)$.
We then calculate the deviation of each cloud from the average for its bin, $\delta n_{\bot,i}(p,T)= n_{\bot,i}(p,T)-\overline{n}_\bot(p,T)$, and calculate the two-body fluctuation correlation function  $\alpha_i(p_{a},p_{b},T)=\delta n_{\bot,i}(p_a,T)\delta n_{\bot,i}(p_b,T)$.
The averages of those quantities, $\overline{\alpha}(p_{a},p_{b},T)$, are displayed in Fig. \ref{fig:alpha}.
Along a line from lower left to upper right is the fluctuation power spectrum $\overline{\alpha}(p,p,T)$.
From upper left to lower right is the opposite-momentum power spectrum  $\overline{\alpha}(p,-p,T)$.
Figure \ref{fig:beta} normalizes the latter by the former, and is a measure of fractional correlations of opposite momenta, $\beta =2\overline{\alpha}(p,-p,T)/[\overline{\alpha}(p,p,T)+\overline{\alpha}(-p,-p,T)]$.
Together these show qualitative changes as a function of temperature and lattice depth.
In concert with our measurement of in-plane coherent fraction, we identify those differences with a set of phases of matter.

Our atomic gases are finite in size and harmonically trapped in all three dimensions.  While disorder  plays a strong role in the onset of superfluidity in each pancake, our harmonic trap favors emergence of order from its center. Consequently there are likely multiple  phases of matter in most of our temperature bins.  The 2.8-kHz lattice exhibits several competing varieties of correlation in $\overline{\alpha}$ as a function of temperature. Isotropic thermal correlations dominate along the lattice between 250 and 175 nK as seen in Figs. \ref{fig:alpha}(a)--(c), when the coherent fraction in plane is insufficient to give rise to interference effects.  The lack of any structure in the normalized correlations $\beta$ is apparent in Fig. \ref{fig:beta}(a). The apparent nonzero correlations in the thermal state derive from the finite width of the temperature bins.  Colder clouds have  smaller rms widths, and vice versa, so fluctuation amplitudes are symmetric about $p_{\bot}=0$ and nonzero. The thermal nature of the correlations imply that $\rho_s^{\bot}=0$. Below 200 nK, we begin to see the growth of a coherent fraction in Fig. \ref{fig:cfrac}, implying that  $\rho_s^{\parallel}>0$. Concurrent $\rho_s^{\bot}=0$ and $\rho_s^{\parallel}>0$ constitute evidence of a sliding phase, a phase exhibiting 2D superfluidity in a 3D bulk, consistent with an anomalous Griffiths phase \cite{VojtaSliding,DemlerSliding,LFSliding}.

Correlations in our 2.8-kHz lattice begin to exhibit structure below 175 nK as seen in Fig. \ref{fig:alpha}(d). We use a multi peak Gaussian fit as a generic localized distribution to analyze our $\beta$ distributions. Strong positive correlations in ($p_{\bot}$,$-p_{\bot}$) emerge around $p_{\bot}=0$, and momenta corresponding near wavelengths of 2.4, 3.0, 4.9, and 10.8 $\mu$m. The 4.9- and 10.8- $\mu$m modes blend together into one broad peak at temperatures below 50 nK, as do those at 2.4 and 3.0 $\mu$m, and at our lowest temperatures 3.0 $\mu$m correlations are suppressed. These emergent length scales are distinct from the periods composing  our disordered potential (1.21, 1.37, 1.58, 2.02, 2.50, 2.79, and 4.70 $\mu$m). Both the momenta peaks in the correlation spectra and the lattice recoil momenta were measured with the same optical system, thus there is no scale-factor uncertainty between the two.
There is a peak in $\beta$ nearby the 2.50-$\mu$m lattice.
However, there is nothing special in our potential about this component, for it is neither the shortest nor the longest period in the potential,
and its amplitude and phase are chosen in the same manner as every other component.
The correlations on the 4.9- and 10.8-$\mu$m length scales are larger than any imposed by our disorder.
We do not see any peaks in $\overline{\alpha}$ corresponding to the recoil momenta of the four lattices with the smallest periods.

While correlation peaks may emerge in completely incoherent lattice systems due to uncorrelated phase noise \cite{FollingFluct,IanFluct}, this does not explain our results.
We confirmed this with a simple 1D single-particle simulation of noninteracting thermal distributions in our disordered lattice, which did not match our results.
The simulations show correlations at low temperature, which depend on choice of relative phases,  exist for each sub-lattice below 50 nK, and emerge first in high momentum states.
In contrast, experimentally observed correlation peaks are evident at higher temperatures ($\sim$175 nK), only occur at a few momenta, and emerge first at low momentum.

The length scales of fluctuations depend on two factors. First, the lattice imposes structure on  the coherent fraction; second, the temperature determines the size of coherently connected lattice sites (``puddles'') and the distances between them. As the temperature is lowered, the system transitions to a fully connected  puddle and lowering the temperature further has no effect.
We attribute this transition to the population of many-body phonon modes as more pancakes undergo the BKT transition, leading to larger superfluid puddles and increasing overall coherent fraction. This is the process described by \cite{DemlerSliding}, as the anomalous Griffiths phase proceeds towards the superfluid transition with decreasing temperature.

This structure first emerges as peaks in Fig. \ref{fig:alpha}(d), and then settles to its final forms in \ref{fig:beta}(a) below 100 nK.
As mentioned, we interpret the settled form of $\beta$ below 100 nK as the establishment of complete connectivity. This suggests that below 100 nK the system has crossed over to a Griffiths superfluid phase from an anomalous Griffiths phase. While we cannot measure phase stiffness directly, we expect it to be quite small in this regime \cite{DemlerSliding,LFSliding}.

The same trend can be seen in the 5.6-kHz lattice data in Fig. \ref{fig:beta}(b), but at much lower temperatures, despite the coherent fraction's concurrent growth with the 2.8-kHz lattice.  The disordered lattice in this data set is twice as deep, so we would expect the most depleted pancake's phase-space density, and thus transition temperature, to drop by $1/e$, and we see no evidence it ever leaves the sliding phase.

\begin{figure}
\includegraphics{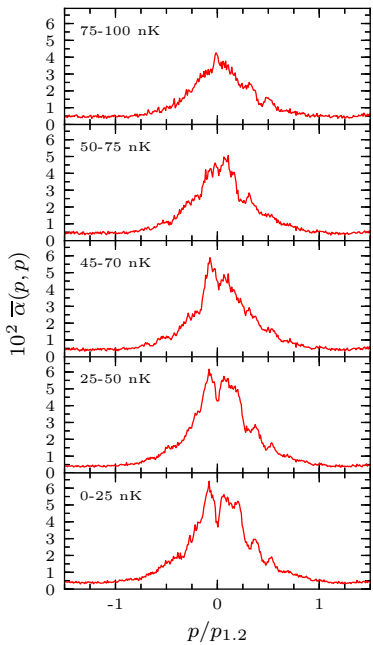}%
\caption{\label{AbsFlucts}Displayed are the square fluctuations $\overline{\alpha}(p,p)$ for a few choices of temperature in our 2.8-kHz lattice. A drop in fluctuations at $p=0$ emerges below 70 $(\pm 2.5)$ nK.}
\end{figure}

At temperatures below 70 nK, in our 2.8-kHz lattice, we observe a drop in the fluctuation power spectrum $\overline{\alpha}(p,p)$ at $p=0$, as shown in Fig. \ref{AbsFlucts}. Its contrast increases with decreasing temperature. We take this as evidence of the suppression of fluctuations due to Bose statistics of a macroscopically occupied, nonlocal state.  This is consistent with the onset of finite phase stiffness along the disorder. We find it difficult to draw a clear line between a superfluid Griffiths phase and a full 3D superfluid. The literature does not discriminate between the two phases using parameters accessible with our system \cite{DemlerSliding,VojtaSliding}, and Ref. \cite{LFSliding} does not identify a distinction. If there are two such distinct phases, this macroscopically ordered region is likely the latter.

\section{Conclusion}

Using our HiBAL we engineered an optical field isotropic in two dimensions and disordered in the third dimension.
We explored the phase diagram of a previously unrealized class of disordered system. The momentum distributions of individual realizations vary wildly, thus we examined noise correlations as our experimental measure.
Although we cannot directly measure phase stiffness (the primary parameter calculated theoretically), we observe trends with temperature consistent with the predictions of this Griffiths system. At high temperatures, distributions show no coherent (low momentum) features in plane and no correlations out of plane.
As the temperature lowers, a low momentum (Thomas-Fermi-like) feature appears in plane, suggesting superfluidity in plane, while the out-of-plane correlations are still absent.
At even lower temperatures, coherence between planes becomes apparent as a $p_\bot=0$ peak in the out-of-plane correlations appears along with correlation features at nonzero momenta. Notably we observe discrete momentum peaks in the correlations unrelated to the momentum components of the disordered lattice.
This suggests local correlations as some regions within the disordered potential become phase coherent, which is characteristic of the anomalous Griffiths phase.
Finally, below 100 nK the noise correlations change little, suggesting full three-dimensional superfluidity (or at least coherence), although the correlation function is still quite different than a superfluid in a lattice without disorder.
At a deeper lattice potential, similar phenomena are observed although the onsets are at lower temperatures, as might be expected as the disorder is better able to prevent coherence buildup.
Direct comparison with theory is needed to fully understand this disordered system, and given experimental limitations, would require calculations beyond phase stiffness.

Our HiBAL is a flexible platform capable of generating arbitrary sets of optical lattices over two spatial octaves with phase, amplitude, and wave-vector control at MHz frequencies. It will enable a large set of experiments, from transport measurements in disordered systems and the production of Hamiltonians for the study of Floquet physics, simultaneous with Bragg spectroscopy of all of the above.

\begin{acknowledgments}
We would like to thank I. Spielman for the extensive discussions during the preparation of this manuscript.
\end{acknowledgments}

\end{document}